\documentclass[aps,prd,twocolumn,superscriptaddress,preprintnumbers,amsmath,amssymb]{revtex4-2}

\usepackage{booktabs}

\usepackage[T1]{fontenc}

\usepackage{epsfig,slashed,nicefrac}
\usepackage[utf8]{inputenc}
\usepackage{color}
\usepackage{mathtools} 
\usepackage{hyperref}
\usepackage{float}
\usepackage{eqnarray,amsmath}
\usepackage{color}
\usepackage{newtxtext,newtxmath}
\usepackage{soul}
\usepackage[letterpaper,margin=1in]{geometry}
\usepackage{url}
\usepackage{lineno}
\usepackage{times}
\usepackage{multirow}
\usepackage{longtable}

\begin{document}


\title[Toponium: the smallest bound state and simplest hadron in quantum mechanics]{Toponium: the smallest bound state and simplest hadron in quantum mechanics}

\author{Jing-Hang~Fu}
\thanks{These authors contributed equally to this work.}
\affiliation{School of Physics, Beihang University, Beijing 100083, China.}

\author{Yu-Ji~Li}
\thanks{These authors contributed equally to this work.}
\affiliation{Institute of Modern Physics, Fudan University, Shanghai 200433, China.}

\author{Hui-Min~Yang}
\affiliation{Center of High Energy Physics, Peking University, Beijing 100871, China.}

\author{Yu-Bo~Li}
\affiliation{School of Physics, Xi‘an Jiaotong University, Xi‘an 710049, China.}

\author{Yu-Jie~Zhang}
\thanks{zyj@buaa.edu.cn (Corresponding author)}
\affiliation{School of Physics, Beihang University, Beijing 100083, China.}
\affiliation{Peng Huanwu Collaborative Center for Research and Education, Beihang University, Beijing 100191, China.}

\author{Cheng-Ping~Shen}
\thanks{shencp@fudan.edu.cn (Corresponding author)}
\affiliation{Key Laboratory of Nuclear Physics and Ion-beam Application (MOE) and Institute of Modern Physics, Fudan University, Shanghai 200443, China.}
\affiliation{School of Physics, Zhengzhou University, Zhengzhou 450001, China}

\begin{abstract}
We explore toponium, the smallest known quantum bound state of a top quark and its antiparticle, bound by the strong force. With a Bohr radius of $8\times 10^{-18}$~m and a lifetime of $2.5 \times 10^{-25}$~s, toponium uniquely probes microphysics. Unlike all other hadrons, it is governed by ultraviolet freedom. This distinction offers novel insights into quantum chromodynamics. Our analysis reveals a toponium signal exceeding $5\sigma$ in the distribution of the cross section ratio between $e^+e^- \rightarrow b\bar{b}$ and $e^+e^- \rightarrow q\bar{q}$ ($q=b,c,s,d,u$), based on 400~fb$^{-1}$ of data collected at $\sqrt{s}\approx 341~{\rm GeV}$. This discovery enables a top quark mass measurement with an uncertainty reduced by a factor of ten compared to current precision levels. Moreover, this method improves the systematic uncertainty by at least a factor of 2.7 compared to any other possible methods.
\end{abstract}

\maketitle

\textit{Introduction}---The top quark ($t$), the heaviest known element particle, plays a pivotal role in Standard Model (SM) of particle physics~\cite{D0:2004rvt,ATLAS:2024dxp}. Its unique properties, particularly its mass, 
have far-reaching implications for our understanding of fundamental forces and the stability of the universe. Toponium, a bound state of a top quark and an anti-top quark ($\bar{t}$), represents the smallest quantum mechanical bound state known to date. With the smallest Bohr radius of $8\times 10^{-18}$~m and an extremely short lifetime of $2.5 \times 10^{-25}$ s, the study of toponium offers a unique opportunity to explore fundamental aspects of quantum mechanics~\cite{Maltoni:2024csn,ATLAS:2023fsd,Aguilar-Saavedra:2024fig,han2024measuringquantumdiscordlhc}, quantum gravity~\cite{Domenech:2020yjf}, and extra dimensions~\cite{Giudice:1998ck} at scales of $10^{-18}$ m and $10^{-25}$ s.

The top quark mass ($m_t$) is a crucial fundamental parameter in SM, as $m_t$ significantly influences our understanding of SM, particularly its vacuum stability, and provides a window into potential new physics beyond SM~\cite{ATLAS:2022vkf,CMS:2022dwd,Alekhin:2012py,Branchina:2014rva,Hall:1993gn,Denner:1989xb}. Furthermore, these precise $m_t$ determinations have far-reaching implications for cosmology~\cite{RodriguezRoman:2018tri}, and the source of inflation in the universe~\cite{Masina:2011un}.

The most precise measurement to date~\cite{ATLAS:2024dxp} yields an on-shell mass of $m_t^{\text{OS}} = (172.52 \pm 0.14_{\rm stat.} \pm 0.30_{\rm syst.} )~{\rm GeV}= (172.52 \pm 0.33)~{\rm GeV}$. While the statistical uncertainty (${\rm stat.}$) can be reduced with longer data acquisition time, the systematic uncertainty (${\rm syst.}$) defines the ultimate precision of the measurement.
The most precise method for determining $m_t$ involves the near-threshold production of $t\bar{t}$ at future electron positron ($e^+e^-$) colliders, such as the Circular Electron Positron Collider (CEPC)~\cite{thecepcstudygroup2018cepcconceptualdesignreport,CEPCStudyGroup:2023quu} and the Future Circular Lepton Collider (FCC-ee)~\cite{FCC:2018evy}. Projections suggest that $m_t$ could be measured to a precision of $25-59$~MeV at CEPC~\cite{Li:2022iav} and $40-75$~MeV at FCC-ee~\cite{schwienhorst2022reporttopicalgroupquark}, with statistical uncertainties as low as 9~MeV at integrated luminosities of $100-200$~fb$^{-1}$ around 343 GeV, although systematic uncertainties will dominate. These uncertainties arise from multiple sources, such as beam energy calibration, theoretical modeling, and detector performance~\cite{Seidel:2013sqa,Li:2022iav}. Therefore, the accurate determination of $m_t$ relies heavily on the development of novel method and advanced systematic control.

In addition to the on-shell mass $m_t^{\text{OS}}$ and other definitions, the 1S mass of top quark, defined as $m_t^{\text{1S}} = m_{J_t}/2$ (where $J_t$ denotes the spin-triplet ground state of toponium)~\cite{Hoang:2013uda}, constitutes the sole scheme- and model-independent measure of the top quark mass $m_t$~\cite{Marquard:2016dcn,Corcella:2019tgt,Hoang:2020iah}. 
Furthermore, toponium distinguishes itself among hadrons because its behavior is governed by ultraviolet freedom, rather than infrared slavery of quantum chromodynamics (QCD)~\cite{Shepherd:2016dni,Ball:2022qks}. This unique characteristic enables toponium to be theoretically described by perturbative QCD (PQCD), thereby offering new insights into confinement within QCD~\cite{Shepherd:2016dni,Witten:1980ez,Guan:2024ccw,Liu:2024fly}.

In this article, we propose a method to discover $J_t$ and accurately determine $m_t^{\text{1S}}$ by measuring the cross section ratio:
\begin{equation}
\label{eq:rb}
R_b = \frac{\sigma_{\rm Born}(e^+e^- \rightarrow b \bar{b})}{\sum_{q=u,d,s,c,b}\sigma_{\rm Born}(e^+e^- \rightarrow q \bar{q})}
\end{equation}
at $e^+e^-$ colliders near $2m_t^{\text{1S}}$, where $b$ is the bottom quark and $q$ is a quark for $u$,~$d$,~$s$,~$c$, or $b$. Future measurements of $R_b$ at the $Z$ pole are expected to achieve precision levels of $4 \times 10^{-5}$ at CEPC~\cite{thecepcstudygroup2018cepcconceptualdesignreport} and less than $6 \times 10^{-5}$ at FCC-ee~\cite{Blondel:2021ema}. The relative precision of $R_b$ at $365$~GeV with $1.5~{\rm ab}^{-1}$ data will be $9.6\times 10^{-4}$ at FCC-ee~\cite{greljo2024newphysicsflavortagging}, which results an absolute precision level of $1.4\times 10^{-4}$. The other possible methods to identifying toponium can be found in Refs.~\cite{Fuks:2021xje,Aguilar-Saavedra:2024mnm,jiang2024studytoponiumspectrumassociated} and related papers.

\textit{Modeling $J_t$ Interaction}---We introduce the Lagrangian density $\mathcal{L}_{J_t}$ of the field $J_t$ into that of SM, $\mathcal{L}_{\text{SM}}$~\cite{Fuks:2021xje} as follows:
\begin{eqnarray}
\label{eq:Lagrangian}
\mathcal{L}_{J_t} &=& \sum_{f=b,e} \bar{f} \left( g^V_{{J_t}ff} \gamma_\mu + g^A_{{J_t}ff} \gamma_\mu \gamma^5 \right) f {J_t}^\mu + \ldots,
\end{eqnarray}
where the coupling factors $g^V_{{J_t}ff}$ and $g^A_{{J_t}ff}$ are determined by non-relativistic QCD (NRQCD)~\cite{Bodwin:1994jh}, $f$ represents the field of either the $b$ quark or the electron, and ${J_t}^\mu$ is the field of $J_t$.

The ${J_t}$ bound state results from the Schr\"{o}dinger equation with a potential between $t$ and $\bar{t}$, which is simplified to a Coulomb plus linear potential based on lattice QCD results~\cite{Kawanai:2013aca,Koma:2006fw}:
\begin{eqnarray}
\label{eq:potential}
V(r) = -\frac{\lambda}{r} +\sigma r, 
\end{eqnarray}
where $r$ is the distance between $t$ and $\bar{t}$, the Coulombic coefficient is $\lambda=0.285\pm 0.011$ at an infinite quark mass, and the string tension is $\sigma=0.206~{\rm GeV}^2$~\cite{Kawanai:2013aca,Koma:2006fw}.
The Bohr radius of toponium is given by $r_B = {2}/{(m_t^{\text{OS}}\lambda)}$, which evaluates to $ {1}/24.584~\rm{GeV}^{-1}$. The $ \sigma r$ term in the potential can be neglected, as its contribution to the binding energy of $J_t$ is $-\langle \sigma r \rangle = -{3 \sigma r_B}/{2} = -0.013~{\rm GeV}$. For the finite top quark mass, we will get a larger $\lambda$.

Additionally, the result from the SU(3) lattice gauge theory suggests a slightly higher value of $\lambda \sim 0.293$~\cite{Luscher:2002qv}.
On the other hand, the production of $\eta_t$ (spin-singlet S-wave toponium) at 13 TeV reported by CMS is $7.1\pm 0.8$~pb in Ref.~\cite{Jeppe:2024uki}. Assuming identical masses and wave functions for the $\eta_t({\rm 1S})$ and $J_t({\rm 1S})$, we find the production of $\eta_t({\rm 1S})$ at 13~TeV at LHC is 6.24 pb with $\lambda=0.285$. This estimate agrees with the value of 6.43 pb reported in Refs.~\cite{Fuks:2021xje,Aguilar-Saavedra:2024mnm}. And summing higher $\eta_t({\rm nS})$ states results in a total production of $\eta_t({\rm nS})$ at 13 TeV at LHC is 6.24~pb$\times \sum 1/n^3 =7.49$~pb.
Recent experiments indicate that the coefficient $\lambda = 0.309 \pm 0.010$~\cite{Fu:2025zxb} is obtained from the larger $\eta_t$ cross-section measurement of $8.8^{+1.2}_{-1.4}$ pb at $\sqrt{s}=13~{\rm TeV}$ measured by CMS~\cite{CMS:2025kzt}. 
In this work, we calculate the systematic uncertainties of $m_t^{\rm 1S}$ from $\lambda$ using the conservative parameter range $\lambda \in [0.285 - 0.011,\, 0.309 + 0.010] = [0.274,\, 0.319]$.

We derive the binding energy $B_{J_t}$, mass $ m_{J_t} = 2m_t^{\text{OS}} - B_{J_t} = 2m_t^{\text{1S}} $, and the square of the wave function at the origin $|\psi_{J_t}(0)|^2$:
\begin{eqnarray}
\label{eq:bmPsi}
B_{J_t} &=& \frac{\lambda^2 m_t^{\text{OS}}}{4} = (3.503 \pm 0.270)~{\rm GeV}, \nonumber \\
m_{J_t} &=& (341.537 \pm 0.653_{m_t^{\text{OS}}} \pm 0.270_{\lambda})~{\rm GeV}\nonumber \\
&=& (341.537 \pm 0.707)~{\rm GeV}, \nonumber \\
|\psi_{J_t}(0)|^2 &=& \frac{(\lambda m_t^{\text{OS}})^3}{8\pi}.
\end{eqnarray}
The uncertainty in $ m_{J_t} $ is mainly due to $ m_t^{\text{OS}} $. The Bohr radius of toponium is calculated to be $8\times 10^{-18}$ m.

The $ B_{J_t} $ can also be calculated in PQCD, yielding $(1.57_{\rm LO}+0.65_{\rm NLO}+0.49_{\rm NNLO}+0.25_{\rm NNNLO})~{\rm GeV}= 2.96 $ GeV at next-to-next-to-next-to-leading order (NNNLO) with a renormalization scale of $ \mu = 32.6 $ GeV for $ m_{J_t} $ and $ \mu = 30 $ GeV for $ m_t^{\text{OS}} $ \cite{Beneke:2005hg}. With $m_t^{\text{OS}} = 172.52~{\rm GeV}$~\cite{ATLAS:2024dxp}, the $\tt{QQbar\_threshold}$ package~\cite{Beneke:2016kkb,Beneke:2017rdn,beneke2024thirdordercorrectiontopquarkpair} yields: 
\begin{eqnarray}
\label{eq:bindEnergyToN3LO}
B_{J_t}^{\rm PQCD} =&(1.680_{\rm LO}+0.657_{\rm NLO}+0.504_{\rm NNLO}\nonumber\\
&+0.243_{\rm NNNLO})~{\rm GeV}= 3.084~{\rm GeV}
\end{eqnarray}
at NNNLO with $ \mu =1/r_B= 24.584$~GeV, where $r_B$ is the Bohr radius. However, the convergence of the perturbative series up to NNNLO is still unsatisfactory, necessitating consideration of higher order corrections. Including the Higgs and $\gamma$ contributions~\cite{Beneke:2015lwa}, the $B_{J_t}^{\rm PQCD}$ is 3.194 GeV, which is comparable to the $B_{J_t}$ in Eq.(\ref{eq:bmPsi}).

The running parameters are chosen at the energy scale of $m_{J_t}$ for both theoretical calculations and experimental simulations~\cite{ParticleDataGroup:2024cfk,ATLAS:2024dxp,Proceedings:2019vxr,Herren:2017osy}:
\begin{eqnarray}
\label{eq:parameters}
m_b=2.568(10)~{\rm GeV}&,&~m_c=0.575(3)~{\rm GeV},\nonumber\\
m_H=125.20(11)~{\rm GeV}&,&~\alpha_s=0.09844(62),\nonumber\\
m_Z=91.1880(20)~{\rm GeV}&,&~\alpha=1/126.04(1),\nonumber\\
m_W=80.3692(133)~{\rm GeV}&,&~\cos \theta_{\rm W}=m_W/m_Z.
\end{eqnarray}

The width $\Gamma_{J_t} $ of $ J_t $ is dominated by twice the width of the top quark $\Gamma_{t}$~\cite{chen2023topquarkdecaynexttonexttonexttoleadingorder,Chen:2023dsi}:
\begin{eqnarray}
\Gamma_{J_t} &=& 2\Gamma_t \left( 1 - \frac{\lambda^2}{8} \right) + \Gamma_{\rm Anni}
= (2.603 \pm 0.019)~{\rm GeV},
\end{eqnarray}
where the coefficient $ \left( 1 - \lambda^2/8\right)$ comes from corrections for the fact that bound top quarks are different from free top quarks~\cite{Uberall:1960zz}. The annihilation decay width of $J_t$, $\Gamma_{\rm Anni}=0.0093~{\rm GeV}$, is calculated using NRQCD \cite{Bodwin:1994jh}.

\begin{figure}[htbp!]
\centerline{\includegraphics[width=0.475\textwidth]{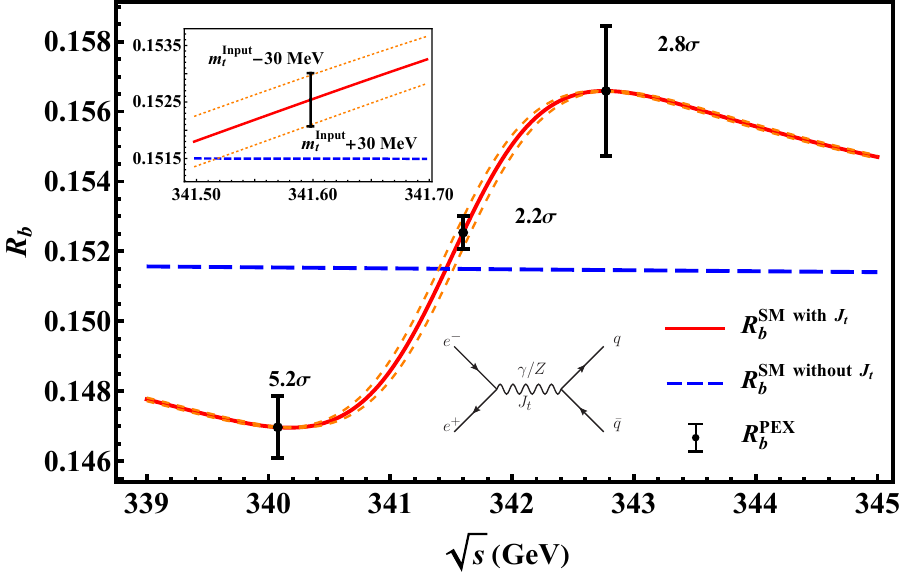}}
\caption{\label{fig:Rb}The blue dashed line represents $R_b$ distribution in SM without $J_t$, the red solid line represents $R_b$ distribution in SM with $J_t$, and the statistical significance of the three energy points is indicated. The upper left insert shows $R_b$ values for $m_t^{\rm Input}$ and $m_t^{\rm Input} \pm 30~\text{MeV}$ near the second energy point. The lower right insert shows the LO Feynman diagram for $e^+e^- \rightarrow q \bar{q}$, where the propagator can be a photon, a $Z$, or a $J_t$.}
\end{figure}

With the Breit-Wigner formula~\cite{Fu:2023uzr}, the Feynman amplitude of the $J_t$ medium particle is:
\begin{eqnarray}
{\mathcal{M}}( J_t)\propto\frac{\lambda^3}{s-m_{J_t}^2+i \Gamma_{J_t}m_{J_t}},
\end{eqnarray}
where $\sqrt{s}$ is the center of mass energy. The real part of the denominator $s-m_{J_t}^2$ goes from negative to 0 to positive as $\sqrt{s}$ goes over $m_{J_t}$. In the amplitudes ${\mathcal{M}}(\gamma)$ and ${\mathcal{M}}(Z)$ of the photon and the $Z$ boson in SM respectively, the sign of the term $s - m^2$ remains unaltered. This results in a characteristic dip-bump structure in the $R_b$ distribution, as shown in Fig.~\ref{fig:Rb}, while the $R_b$ distribution of SM without $J_t$ is very close to a horizontal straight line~\cite{ParticleDataGroup:2024cfk,Djouadi:1989uk,Chetyrkin:1997qi,Novikov:1999af,Chen:2022vzo,Chetyrkin:2000zk}:
\begin{eqnarray}
\label{eq:RbNNNLO}
R_{b}^{\rm SM} &=&\bigg[151,\!550.2_{\rm LO}-50.6_{\rm NLO}+0.8_{\rm NNLO}+0.0_{\rm NNNLO}\nonumber\\
&&\hspace{-0.8
cm}-27.4 \frac{\sqrt{s}-m_{J_t}}{\rm GeV}+0.1\left( \frac{\sqrt{s}-m_{J_t}}{\rm GeV}\right)^2+...\bigg]\times 10^{-6}.
\end{eqnarray}
By observing the deviation of $R_b$ distribution from the straight line, we can identify $J_t$, and a precise measurement of this deviation allows us to determine $m_t^{\text{1S}}$.

We choose the central values of $m_{J_t}$ and $\lambda$ as input, and our results will demonstrate the validity of this method within the corresponding ranges. Since $\Gamma_{J_t}$ can be predicted within a certain range in SM without affecting our conclusions, we fix $\Gamma_{J_t}$ to its central value. The experimental cross section $\sigma$ is given by the following expression~\cite{Fu:2023uzr,Beneke:2017rdn}:
\begin{eqnarray}
\label{eq:EXCrossSection}
\sigma(\sqrt{s},\delta_{\sqrt{s}})=\int d\sqrt{s^{\prime}}\frac{1}{\sqrt{2\pi}\delta_{\sqrt{s}}}e^{-\frac{(\sqrt{s}-\sqrt{s^{\prime}})^2}{2\delta_{\sqrt{s}}^2}}\nonumber\\
\times\int dx~F(x,\sqrt{s^{\prime}})\bar{\sigma}(\sqrt{xs^{\prime}})=f_{\rm ISR}~\sigma_{\rm Born}.
\end{eqnarray}
In this formula, $\delta_{\sqrt{s}}$ is the center of mass energy spread, and the initial state radiation (ISR) factor $F(x,\sqrt{s^{\prime}})$ is given in Refs.~\cite{Kuraev:1985hb,Beneke:2017rdn,Ruiz-Femenia:2001qlg}. Then, we can obtain $f_{\rm ISR}$, the factor that includes the effects of energy spread and ISR.

\textit{Experimental Simulation}---In this article, Monte Carlo (MC) simulations of $e^+e^-$ collisions have been performed. The physics events are generated using $\tt Whizard$~\cite{Kilian:2007gr}, where the effects of ISR and energy spread have been taken into account. Subsequently, $\tt Pythia6$~\cite{Sjostrand:2006za} is applied for parton showering, hadronization, and the decay processes of the short-life hadrons. Finally, $\tt Delphes$~\cite{deFavereau:2013fsa}, configured with the CEPC detector conditions~\cite{thecepcstudygroup2018cepcconceptualdesignreport}, is used to simulate the detector response. To measure $R_b$, samples with 2 fermions ($u\bar{u},d\bar{d},s\bar{s},c\bar{c},b\bar{b}$) are generated. Moreover, due to limitations in detector performance, some jets with low energy or those produced at extreme angles relative to the detector's central axis~\cite{thecepcstudygroup2018cepcconceptualdesignreport,FCC:2018evy} may go undetected. To address this, several multi-jet backgrounds from $WW$, $ZZ$, and $ZH$ decays to four-quark final states are simulated. In order to enhance the sensitivity for $J_t$ signal, the event selection criteria are optimized as discussed below.

Jets from $q\bar{q}$ are clustered from all the well reconstructed tracks with an anti-kt algorithm~\cite{Cacciari:2008gp}. To select high-quality jets, we require jet transverse momentum ($p_t$) greater than 15 GeV. To include the case of gluon final state radiation in $q\bar{q}$ final state, the selected events are supposed to have 2 or 3 jets. As the detector does not cover all directions, the ISR photons and jets close to the beam direction may escape detection. Thus, the effective invariant mass of the final states $\sqrt{s^{\prime}}$ needs to be considered. We require $\sqrt{s^\prime}> \sqrt{s}-10~{\rm GeV}$ here~\cite{OPAL:1996vpu,ALEPH:2006jhv,Beneke:2017rdn}. For the measurement of $R_b$, only events with at least two $b$-tagged jets~\cite{Wu:2018awx} are categorized as $b\bar{b}$ events, while the remaining events are considered as light quark pair processes. For the selected signal events, we require the invariant mass of jet pairs $M_{bb}$ or $M_{j_{1}j_{2}}$ greater than 150 GeV to suppress the backgrounds from $W/Z$ hadronic decays, where ${j_{1}j_{2}}$ denote the two jets with the highest and the second-highest energies.
Events passing all of the criteria are recognized as signal events, and the detection efficiencies of different processes in the three energy points are shown in Table~\ref{tab:SigAndEfficiency}. 

\begin{table*}[hbtp!]
\caption{\label{tab:SigAndEfficiency}The experimental cross sections $\sigma(\sqrt{s})$ (in fb), along with the detection efficiencies $\varepsilon_{\text{tag}}$ (in \%, for double $b$-tagged events) and $\varepsilon_{\text{hadron}}$ (in \%, for hadronic events), are provided for various processes in the $R_b$ measurements. Here, $jj$ denotes $u\bar{u}$, $d\bar{d}$, $s\bar{s}$, $c\bar{c}$, or $gg$.}
\centering
\begin{tabular}{crrrrr}
\toprule
Process & $\sigma(340.076)$ (fb) &~~~$\sigma(341.598)$ (fb) &~~~$\sigma(342.771)$ (fb) &~~$\varepsilon_{\text{tag}}(\%)$&~~$\varepsilon_{\text{hadron}}(\%)$ \\
\midrule
$b\bar{b}$ & 866.99 & 880.50 & 896.87 &46.42 & 52.70 \\
$jj$ & 4,977.58 & 4,945.68 & 4,927.87 &0.04 &52.48 \\
$ZZ/ZH \to b\bar{b}b\bar{b}$   & $16.17/12.14$ & $16.01/11.88$ & $15.86/11.70$  & 0.31 & 0.31 \\
$ZZ/ZH \to b\bar{b}jj $ & $116.42/46.06$ & $115.11/45.11$  & $114.19/44.42$  & 1.64 & 2.21 \\
$ZZ/ZH/WW \to jjjj $  &~~$209.21/8.60/5,\!263.55$ &~~$207.11/8.42/5,\!202.29$ &~~$205.38/8.29/5,\!157.25$ & 0.00 & 5.19 \\
\bottomrule
\end{tabular}
\end{table*}

\textit{Measuring $R_b$}---
The $R_b$, an observable which can be precisely measured in a pseudo-experiment with the CEPC scenario, is obtained by~\cite{ALEPH:2006jhv}:
\begin{eqnarray}
\label{eq:rbDefine}
R_b=\frac{N_{\text{tag}} F_{\text{hadron}} ^{udsc}-N_{\text{hadron}}
F_{\text{tag}}^{{udsc}}}{
N_{\text{tag}}(F_{\text{hadron}}^{udsc}-F^b_{\text{hadron}})+N_{\text{hadron}} (F_{\text{tag}}^b-F_{\text{tag}}^{udsc})}.
\end{eqnarray}
Here, $F=\varepsilon f_{\rm ISR}$, the detection efficiency for $b$ quarks after event selection criteria is denoted by $\varepsilon^{b}_{\text{hadron}}$ (for hadronic events) and $\varepsilon^{b}_{\text{tag}}$ (for double $b$-tagged events). Similarly, the averaged detection efficiency for light quarks ($q = u, d, s, c$) is denoted by $\varepsilon_{\text{hadron}}^{udsc}$ (for hadronic events) and $\varepsilon_{\text{tag}}^{udsc}$ (for double $b$-tagged events), weighted by the cross section of $e^+e^- \to q \bar{q}$. The values are $\varepsilon_{\text{hadron}}^{udsc} = 52.48\%$, $\varepsilon_{\text{hadron}}^{b} = 52.70\%$, $\varepsilon_{\text{tag}}^b = 46.42\%$, and $\varepsilon_{\text{tag}}^{udsc} = 0.04\%$. The correction factors $f^{udsc}_{\rm ISR}$ and $f^b_{\rm ISR}$ are used to model the effects of ISR and energy spread. Meanwhile, $N_{\text{tag}}$ $(N_{\rm hadron})$ is the number of selected double $b$-tagged hadronic events (the number of hadronic events) from the experiment datasets, and can be obtained with information of the cross section, integrated luminosity, and detection efficiency at an energy point $\sqrt{s_i}$:
\begin{eqnarray}
N_{\rm hadron}(\sqrt{s_i})&=&\mathcal{L}_i \sum_{q=udscb}
\varepsilon^{q}_{\rm hadron}\sigma(e^+e^-\to q\bar{q}),\nonumber\\
N_{\text{tag}}(\sqrt{s_i})&=&\mathcal{L}_i \sum_{q=udscb}
 \varepsilon_{\text{tag}}^{q}
\sigma(e^+e^-\to q\bar{q}).
\end{eqnarray}

In pseudo-experiments, we perform an analysis on datasets generated by SM with $J_t$ model to extract $R_b$ value ($R_b^{\text{PEX}}$) at a chosen energy point ($\sqrt{s_i}$). The chi-square statistic $\chi^2$ is then defined as follows:
\begin{eqnarray}
\label{eq:Chi2}
\chi^2_{\rm Model}=\sum_{i}{\left(\frac{R_b^{\text{PEX}}(\mathcal{L}_i,\sqrt{s_i})-R_b^{\rm Model}(\mathcal{L}_i,\sqrt{s_i})}{\Delta R_b^{\text{PEX}}(\mathcal{L}_i,\sqrt{s_i})}\right)^2}.
\end{eqnarray}
The $\mathcal{L}_i$ is the integrated luminosity at $\sqrt{s_i}$, and ${\Delta R_b^{\text{PEX}}(\mathcal{L}_i,\sqrt{s_i})}$ is the uncertainty of $R_b^{\text{PEX}}(\mathcal{L}_i,\sqrt{s_i})$. The fitting function $R_b^{\rm Model}(\mathcal{L}_i,\sqrt{s_i})$ gives a theoretical $R_b$ value at an energy point, where free parameters $m_{J_t}$ and $\lambda$ can be varied to minimize $\chi^2_{\rm Model}$.

Three sets of energy points are selected as $\sqrt{s}=(340.076,341.598,342.771)~{\rm GeV}$, and the corresponding luminosities are determined as $\mathcal{L}=(80,300,20)~{\rm fb}^{-1}$, to make the significance of the $J_t$ signal exceeding $5\sigma$ discussed below. With this statistical strategy, we can obtain the maximum values of $\chi_i^2/\mathcal{L}_i$ (or $\partial R_b/\partial \sqrt{s}=0$) for the first and third energy points, and a $\chi_i^2/\mathcal{L}_i$ value close to zero (but $\partial R_b/\partial m_t$ is maximum) for the second point, used to measure the $m_t^{\text{1S}}$. The significance of the $J_t$ signal is calculated by $\Delta\chi^2/{\rm dof}=(\chi^2_{\text{SM without}~J_t}-\chi^2_{\text{SM with}~J_t}) /{\rm dof}$, where dof is the number of degrees of freedom. The values of $R_b^{\text{PEX}}$ along with their respective statistical uncertainties are shown in Fig.~\ref{fig:Rb}. Considering only the statistical uncertainty, the significance of the $J_t$ signal amounts to 6.26$\sigma$.

\begin{figure}[htbp!]
\centerline{\includegraphics[width=0.475\textwidth]{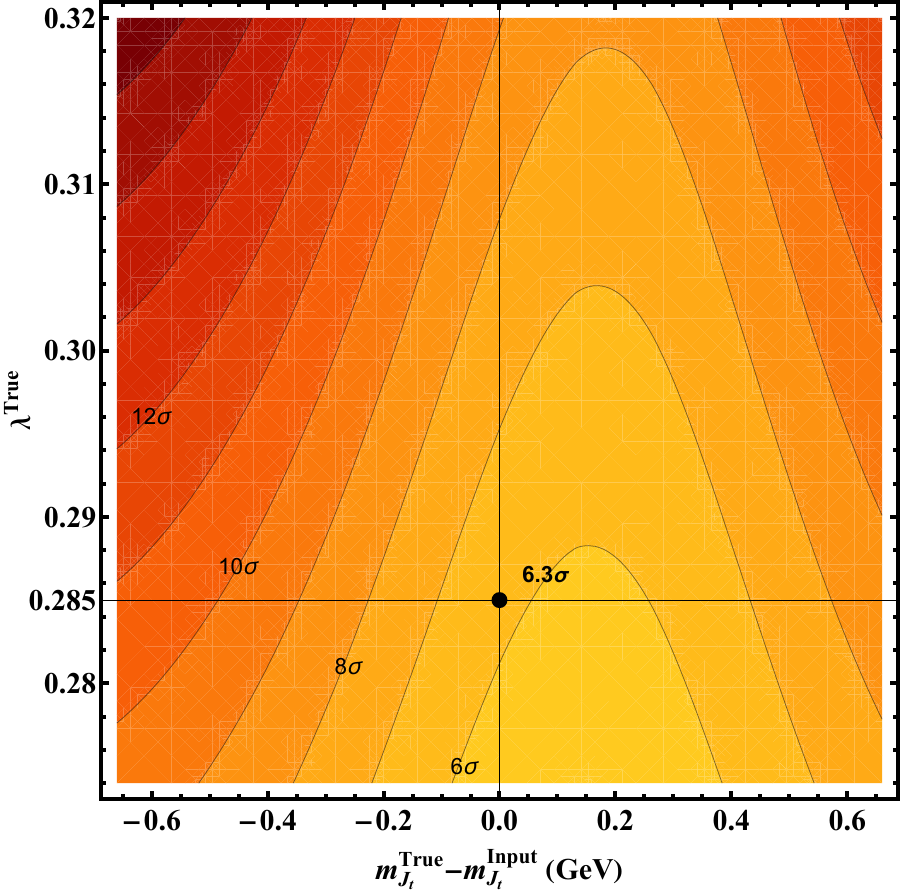}}
\caption{\label{fig:Significance_MJt_lambda}Significance of the $J_t$ signal varies with $m_{J_t}$ within the $1\sigma$ range and $\lambda$ within the $[0.274,\, 0.319]$ range.
}
\end{figure}

The main sources of systematic uncertainties in the $R_b$ measurements are the energy scale of $\sqrt{s}$, energy spread, detection efficiency, and backgrounds, and see Appendix \ref{AppendixA} for details. Which lead to a $J_t$ signal significance of exceeding $5\sigma$ even when the total systematic uncertainties are taken into account. 

When the true value of $m_{J_t}$ varies within $1\sigma$ and $\lambda$ varies within the $[0.274,\, 0.319]$ range, we determine the corresponding significance of the $J_t$ signal. This significance consistently exceeds the $5\sigma$ threshold and is shown in Fig.~\ref{fig:Significance_MJt_lambda}.

\textit{Extraction of $m_t^{\text{1S}}$ and Uncertainties Estimation}---The top quark mass $m_t^{\text{1S}}$ (along with the coefficient $\lambda$) is extracted by minimizing $\chi^2_{\text{SM with}~J_t}$. The statistical uncertainty $({\Delta m_t^{\text{1S}}})_{\rm stat.}$ is found to be {$\rm 32.2~MeV$}, primarily determined by the second energy point $\sqrt{s_2}$. The systematic uncertainties, $({\Delta m_t^{\text{1S}}})_{\rm syst.}$, arise from several sources, listed in Table~\ref{tab:Uncertainties}.

The uncertainty of $\sqrt{s_2}$ energy scale is estimated to be 2.0~MeV at $2m_t^{\text{1S}}$~\cite{Blondel:2019jmp}, which directly enters into the $m_{J_t}$ and contributes 1.0~MeV to $\Delta m_t^{\text{1S}}$. The center of mass energy spread is 458.4 MeV at $\sqrt{s_2}$~\cite{Li:2022iav,CEPCStudyGroup:2023quu}, with an estimated uncertainty of 10\%~\cite{Li:2022iav,Blondel:2019jmp}, contributing an additional {0.2}~MeV to ${\Delta m_t^{\text{1S}}}$. The detection efficiency related uncertainty in $R_b$ is ${ \varepsilon_{\text{hadron}}^{udsc}}/{ \varepsilon_{\text{tag}}^b}$, which is consistent that in the $R_b$ measurement at the $Z$ pole. After including uncertainties from MC statistics and energy-momentum requirement, the uncertainty associated with the detection efficiency in $R_b$ is $19.1\times 10^{-6}$. This is about half of the total uncertainty of the $R_b$ measurement at the $Z$ pole at CEPC, which is $4\times 10^{-5}$~\cite{ thecepcstudygroup2018cepcconceptualdesignreport}. Consequently, this contributes 1.3 MeV to $\Delta m_t^{\text{1S}}$.
By comparing the background contributions estimated in theory and observed in experiment, a relative precision of $(\alpha_s/\pi)^2 \sim 10^{-3}$ can be achieved, contributing {0.9} MeV to $\Delta m_t^{\text{1S}}$. Additionally, $\sin^2\theta_{\rm W}$ can attain an accuracy of $1\times10^{-5}$~\cite{thecepcstudygroup2018cepcconceptualdesignreport}, thereby contributing {0.4} MeV to $\Delta m_t^{\text{1S}}$. The NNLO correction of $R_b$ yields a contribution of {0.1} MeV to $\Delta m_t^{\text{1S}}$. 
Using the conservative range $\lambda \in [0.274,\, 0.319]$, we calculate the $\lambda$-induced systematic uncertainty of $m_t^{\rm 1S}$ to be $8.4~\mathrm{MeV}$, while noting that the energy point scheme was optimized specifically for $\lambda = 0.285$.
The contribution of other sources, such as luminosity and backgrounds from two-photon and $WWZ$ processes, is less than 0.2 MeV for $\Delta m_t^{\text{1S}}$.

Assuming these sources are independent, we combine them in quadrature, yielding a total systematic uncertainty of $({\Delta m_t^{\text{1S}}})_{\rm syst.} = {8.6}$~MeV, as summarized in Table~\ref{tab:Uncertainties}.
The systematic uncertainty of our method is greatly improved over the predictions of previous studies~\cite{schwienhorst2022reporttopicalgroupquark,Li:2022iav} due to simpler final states and differences in the physical quantities to be measured. 
Then we get: 
\begin{eqnarray}
\Delta{ m_t^{\text{1S}}}=(\pm 32.2_\text{{stat.}}\pm 8.6_\text{{syst.}})~{\text{MeV}}=\pm 33.3~{\text{MeV}}.
\end{eqnarray}

\begin{table*}[htpb!]
\caption{\label{tab:Uncertainties}Systematic uncertainties of the $m_t$ measurement in MeV. The ``Other" category includes luminosity measurements and backgrounds from two-photon and $WWZ$ processes.} 
\setlength{\tabcolsep}{10pt}
\centering
\begin{tabular}{lccc}
\toprule
$\Delta m_t$ (MeV) & FCC-ee~\cite{schwienhorst2022reporttopicalgroupquark} & CEPC~\cite{Li:2022iav} & This work \\ 
\midrule
Energy scale      & 3 & 2 &  1.0\\
Energy spread     & -- & $3-5$ & 0.2 \\
Efficiency    & -- & $5-45$ & 1.3 \\
Background        & $10-20$ & $4-18$ & 0.9 \\ 
SM parameters ($\theta_{\rm W},\alpha_s$)        & 3.2 & 17 & 0.4 \\   
Higher order corrections            & $40-45$ & $9-26$ & 0.1 \\   
Potential scheme ($\lambda$) &  -- & -- & 8.4 \\
Other            & -- & 10 & $<0.2$ \\   
\midrule
Total syst.       & $40-74$ & $23-58$ & 8.6  \\ 
\bottomrule
\end{tabular}
\end{table*}

If the luminosities are chosen as $\mathcal{L}=(80,900,20)~{\rm fb}^{-1}$ and $\mathcal{L}=(80,1,\!600,20)~{\rm fb}^{-1}$ for CEPC~\cite{CEPCStudyGroup:2023quu} and FCC-ee~\cite{FCC:2018evy} at (340.076, 341.598, 342.771) GeV, respectively, our results show that, 
the statistical uncertainty, $({\Delta m_t^{\text{1S}}})_{\rm stat.}$, is estimated to be 18.6~MeV for CEPC and 13.9~MeV for FCC-ee, or 11.2~MeV when data from both colliders are combined. A detailed comparison of the top quark mass measurement methods at CEPC~\cite{Li:2022iav}, FCC-ee~\cite{schwienhorst2022reporttopicalgroupquark}, and in this work is presented in Table~\ref{tab:mtRbVstt}.

\begin{table*}[htbp!]
\caption{Detailed comparison of the top quark mass measurement methods: final states, signal and background cross sections, and uncertainty estimation at CEPC~\cite{Li:2022iav}, FCC-ee~\cite{schwienhorst2022reporttopicalgroupquark}, and in this work. }
\label{tab:mtRbVstt}
\centering
\begin{tabular}{cccc}
\toprule
Method & CEPC~\cite{Li:2022iav} & 
FCC-ee~\cite{schwienhorst2022reporttopicalgroupquark} & This work  \\
\midrule
Final states & $ b\bar{b}f\bar{f}f\bar{f}$ & $ b\bar{b}f\bar{f}f\bar{f}$ & $ b\bar{b}$,~$ q\bar{q}$ \\ 	
Measured quantity & $\sigma({t\bar{t}})$ & $\sigma({t\bar{t}})$ & $R_b=\sigma({b\bar{b}})/\sigma({q\bar{q}})$ \\
$\sigma_{\rm signal} $ & $\sim 450~$fb~\cite{Seidel:2013sqa} & $\sim 450~$fb~\cite{Seidel:2013sqa} &~~$\sim900$~fb ($b\bar{b}$),~$\sim$ 6,000~fb ($q\bar{q}$)~~\\
Signal over background & {$\sim 2.9$~\cite{Seidel:2013sqa}} & {$\sim 2.9$~\cite{Seidel:2013sqa}} & {$\sim 89$, $\sim 12$}\\
$\sqrt{s}$ (GeV) & $\sim 343~$GeV &$\sim 343~$GeV &$\sim 341~$GeV \\
Integrated luminosity &100~fb$^{-1}$&200~fb$^{-1}$& $400-2,\!700$~fb$^{-1}$\\
$(\Delta m_t)_{\rm stat.}$  & $9$~MeV & $9$~MeV  & $11.2-32.2$~MeV \\
$(\Delta m_t)_{\rm syst.}$  & $23-58$~MeV &~~$40-75$~MeV & 8.6~MeV \\
$(\Delta m_t)_{\rm total}$  &~~~~$25-59$~MeV~~~~ &~~~~$40-75$~MeV~~ & $14.1-33.3$~MeV \\
\bottomrule
\end{tabular}
\end{table*}

\textit{Discussion and Outlook}---In this article, we propose a method for both discovering toponium and precisely measuring the top quark mass $m_t^{\text{1S}}$ using the observable $R_b$ near the energy threshold $2m_t^{\text{1S}}$ in $e^+e^-$ collisions. The significance of the toponium signal exceeds $5\sigma$, including both statistical and systematic uncertainties with the integrated luminosity of 400 $\rm fb^{-1}$ around 341 GeV. The $m_t^{\text{1S}}$ can be measured with a total uncertainty of only 33.3 MeV, including a systematic uncertainty of 8.6 MeV. This enables a top quark mass measurement with an uncertainty reduced by a factor of 10 compared to the current precision~\cite{ATLAS:2024dxp}. Additionally, polarized beams or the well separation of $b$-jet and $\bar{b}$-jet in detection can reduce the integrated luminosities by an order of magnitude.

\textit{Acknowledgments}---We thank Profs. Kuang-Ta Chao, Gang Li, Yan-Qing Ma, Manqi Ruan, Xiaohu Sun, and Shi-Lin Zhu for valuable and helpful discussions. This work is supported in part by National Key Research and Development Program of China under Contract No.~2024YFA1610503, National Natural Science Foundation of China (NSFC) under contract No.~12161141008, No.~12135005, and No.~12405102, the Fundamental Research Funds for the Central Universities Grant No.~4007022302, and the China Postdoctoral Science Foundation under Grant No.~2024M750049.

\vspace{1em}
\appendix
\section{\texorpdfstring{Systematic uncertainties of $R_b$}{Systematic uncertainties of Rb}}\label{AppendixA}
Taking the $R_b$ measurement at the second energy point as an example, systematic uncertainties are discussed below. The uncertainty in the energy scale of $\sqrt{s_2}$ is estimated to be 2 MeV~\cite{Blondel:2019jmp}, contributing $14.6\times 10^{-6}$ to $\Delta R_b$. The energy spread at the second energy point is 458.4 MeV for CEPC, with a 10\% uncertainty~\cite{Li:2022iav,Blondel:2019jmp} contributing an additional $2.7\times 10^{-6}$ to $\Delta R_b$. By increasing the size of simulated events, the statistical error can be reduced to less than $10^{-4}$, resulting in a contribution of $15.4\times 10^{-6}$ to $\Delta R_b$. The precisions of $p_t$ and $\sqrt{s'}$ are controlled to within 0.10 GeV and 0.34 GeV~\cite{ALEPH:2006jhv,Lai:2021rko}, respectively, leading to $\Delta R_b$ values of $9.1\times10^{-6}$ and $6.7\times10^{-6}$. 
The relative uncertainty in $WW$, $ZZ$, and $ZH$ backgrounds is less than $(\alpha_s/\pi)^2 \sim 0.001$, contributing $13.9\times10^{-6}$ to $\Delta R_b$. The uncertainty in $\sin^2\theta_{\rm W}$, achievable at $1\times 10^{-5}$ at CEPC~\cite{thecepcstudygroup2018cepcconceptualdesignreport,Blondel:2019jmp}, contributes $6.0\times10^{-6}$ to $\Delta R_b$. The uncertainty of $m_b$ in Eq.(\ref{eq:parameters}) contributes $0.2\times10^{-6}$ to $\Delta R_b$, while the uncertainty from $\alpha_s$ is included in higher-order corrections. As shown in Eq.(\ref{eq:RbNNNLO}), the NNLO corrections increase $R_b$ by $1.2\times 10^{-6}$. The uncertainty in $\lambda$, estimated using the conservative parameter range $[0.274,\, 0.319]$, contributes $\Delta R_b = 99.2 \times 10^{-6}$. The other sources include luminosity measurements and backgrounds from two-photon and $WWZ$ processes. Their contributions to ${\Delta R_b}$ are small, with a combined estimation of less than $3.0\times 10^{-6}$. All the systematic uncertainties are summarized in Table~\ref{tab:RbUncertainties}.

\begin{table*}[hbtp!]
\caption{Uncertainties of $R_b^{\rm PEX}$ are given in a unit of $10^{-6}$. The ``Other" category includes luminosity measurements and backgrounds from two-photon and $WWZ$ processes.}
\label{tab:RbUncertainties}
\centering
\begin{tabular}{crrr}
\toprule
Source  & {340.076~GeV} &~~~{341.598~GeV} &~~~{342.771~GeV} \\
\midrule
Energy scale
    &   0.5 &  14.6 & 0.1 \\
Energy spread
    & 36.0 & 2.7 &  58.9   \\
MC statistics
    & 14.8  & 15.4 & 15.8 \\
$p_t$ requirement  
    & 8.9  & 9.1 & 9.3 \\
$\sqrt{s^\prime}$ requirement 
    & 6.4  & 6.7 & 7.0 \\
Backgrounds 
    & 13.5  & 13.9 & 14.2 \\
SM parameters
    & 5.0  & 6.0 & 6.8 \\
Higher order corrections
    & 1.2 &  1.2 & 1.2 \\
Potential scheme ($\lambda$)
    & 98.2 & 99.2  & 21.4  \\
Other
    & 3.0 & 3.0 & 3.0 \\
\midrule
${\rm Sum~ of~ syst.}$ &  107.2 & 103.2  & 67.5 \\
${\rm Stat.}$ &  884.4 &  471.2 & 1,859.3 \\
${\rm Total}$  &  890.9   & 482.4  & 1,860.5 \\
\bottomrule
\end{tabular}
\end{table*}

\bibliographystyle{apsrev4-2_Update}

\bibliography{bibliography}

\end{document}